\documentclass[twocolumn,showpacs]{revtex4}[12pt]
\usepackage{graphicx}
\draft
\begin{document}
\begin{sloppypar}
\title[Electron-acoustic solitary pulses and double layers]{Electron-acoustic solitary pulses and double layers in multi-component plasmas}
\author{A. Mannan$^1$\footnote{email: mannan@na.infn.it}, A. A. Mamun$^2$, and P. K. Shukla$^{3,4}$}
\address{$^1$Dipartimento di Matematica e Fisica, Seconda Universit\`{a} degli Studi di Napoli, Via Vivaldi 43, 81100 Caserta, Naples, Italy\\
$^2$Department of Physics, Jahangirnagar University, Savar, Dhaka-1342, Bangladesh\\
$^3$RUB International Chair, International Centre for
Advanced Studies in Physical Sciences, Faculty of Physics \&
Astronomy, Ruhr-University Bochum, D-44780 Bochum, Germany\\
$^4$Department of Mechanical and Aerospace Engineering \& Center for Energy
Research, University of California San Diego, La Jolla, CA 92930, United States of America}
\begin{abstract}
We consider the nonlinear propagation of finite amplitude electron-acoustic waves (EAWs)
in multi-component plasmas composed of two distinct groups of electrons (cold and hot components),
and non-isothermal ions. We use the continuity and momentum equations for cold inertial electrons,
Boltzmann law for inertialess hot electrons, non-isothermal density distribution for hot ions, and
Poisson's equation to derive an energy integral with  a modified Sagdeev potential (MSP) for
nonlinear EAWs. The MSP is analyzed to demonstrate the existence of arbitrary amplitude EA solitary
pulses (EASPs) and EA double layers (EA-DLs). Small amplitude limits have also been considered and
analytical results for EASPs and EA-DLs are presented. The implication of our results to space and
laboratory plasmas is briefly discussed.
\end{abstract}
%\noindent{\it Keywords}:
\maketitle
\section{Introduction}
The idea of the electron-acoustic wave (EAW) has been conceived by
Fried and Gould \cite{Fried1961} during numerical solutions of the
linear electrostatic Vlasov dispersion equation in a uniform
unmagnetized plasma. It is basically an acoustic-type
waves \cite{Watanabe1977}, in which  inertia is provided by
the cold electron mass, and the restoring force comes from
the hot electron thermal pressure. The ions play the role of a
neutralizing background only. The spectrum of the linear EA waves,
unlike that of the well-known Langmuir waves, extends only up to
the cold electron plasma frequency $\omega_{pc}=(4\pi
n_{c0}e^2/m_e)^{1/2}$, where $n_{c0}$ is the unperturbed cold
electron number density, $e$ magnitude of the electron charge,
and $m_e$  the mass of the electron. This upper wave frequency
limit $(\omega\simeq\omega_{pc})$ corresponds to a
short-wavelength EAW and depends on the unperturbed cold
electron number density $n_{c0}$. On the other hand, the
dispersion relation of the linear EAWs in the long-wavelength
limit [in comparison with the hot electron Debye radius
$\lambda_{dh}=(k_BT_h/4\pi n_{h0}e^2)^{1/2}$, where $T_h$ is the
hot electron temperature, $k_B$ is the Boltzmann constant, and
$n_{h0}$ the unperturbed hot electron number density] is
$\omega\simeq kC_e$, where $k$ is the wave number and
$C_e=(n_{c0}k_BT_h /n_{h0}m_e)^{1/2}$  the EA speed
\cite{Gary1985}. Besides the well-known Langmuir and ion-acoustic
waves, they noticed the existence of a heavily damped
acoustic-like solution of the dispersion equation. It was later
shown that in the presence of two distinct groups (cold and hot)
of electrons and immobile ions, one indeed obtains a weakly
damped EAW \cite{Watanabe1977}, the properties of which
significantly differ from those of the Langmuir waves. Gary and
Tokar \cite{Gary1985} performed a parameter survey and found
conditions for the existence of the EAWs. The most important
condition is $T_c \ll T_h$, where $T_c$ $(T_h)$ is the
temperature of cold (hot) electrons. The propagation
characteristics of the EAWs has also been studied by Yu and
Shukla \cite{Yu1983}, Mace and Hellberg
\cite{Mace1990,Mace1993,Mace2001} and Mace \textit{et al.}
\cite{Mace1991}.

Two-electron-temperature plasmas are known to occur both in laboratory experiments \cite{Derfler1969,Henry1972,Sheridan1991,Pickett2008,Kakad2009,Moon2010,Singh2011,Mondal2012}
and in space environments
\cite{Dubouloz1991,Dubouloz1993,Pottelette1999,Berthomier2000,Singh2001}.
The propagation of the EAWs has received a great deal of
renewed interest not only because the two-electron-temperature
plasma is very common in laboratory experiments and in space, but
also because of the potential importance of the EAWs in
interpreting electrostatic component of the broadband
electrostatic noise (BEN) observed in the cusp of the terrestrial
magnetosphere \cite{Tokar1984,Singh2001}, in the geomagnetic tail
\cite{Schriver1989}, in auroral region
\cite{Dubouloz1991,Dubouloz1993,Pottelette1999}, etc.

The EAWs have been used to explain various wave emissions in
different regions of the Earth's magnetosphere
\cite{Dubouloz1991,Dubouloz1993}. It was first applied to
interpret hiss-emissions observed in the polar cusp region in
association with low-energy $(\sim100~eV)$ upward moving electron
beams \cite{Lin1984}. The EAWs were also utilized to interpret
the generation of the BEN emissions detected in the plasma sheath
\cite{Schriver1989}, as well as in the dayside auroral zone
\cite{Dubouloz1991,Dubouloz1993}. Dubouloz \textit{et al.}
\cite{Dubouloz1991} rigorously studied the BEN observed in the
dayside auroral zone and showed that because of the very high
electric field amplitudes $(100~mV/m)$ involved, the nonlinear
effects must play a significant role in the generation of the BEN
in the dayside auroral zone. Dubouloz \textit{et al.}
\cite{Dubouloz1991,Dubouloz1993} also explained the
short-duration $(<1 s)$ burst of the BEN in terms of electron
acoustic solitary waves (EA-SWs): such EA-SWs passing the
satellite would generate electric field spectra. To study the
properties of EA solitary structures, Dubouloz \textit{et al.}
\cite{Dubouloz1991} considered a one-dimensional, unmagnetized
collisionless plasma consisting of cold electrons, Maxwellian hot
electrons, and stationary ions. El-Shewy \cite{El-Shewy2007}
has investigated the propagation of linear and nonlinear EA-SWs
in a plasma containing cold electrons, nonthermal hot electrons,
and stationary ions. The effects of arbitrary amplitude EA-SWs and
electron-acoustic double layers (EA-DLs) in a plasma consisting
of cold electrons, superthermal hot electrons, and stationary
ions has been considered by Sahu \cite{Sahu2010}.
The EA-SWs in a two-electrons-temperature plasma where ions form
stationary charge neutral background has been observed by Dutta \cite{Dutta2011}.
El-Wakil {\it et al.} \cite{El-Wakil2011} considered cold electrons, nonthermal hot electrons,
and stationary ions, and studied the nonlinear properties of EA-SWs by using time-fractional
Korteweg-de Vries (K-dV) equation. On the other hand,  there are  some space, where
the energetic ions are observed, and ion-temperature can very high, even higher than
electron-temperature \cite{Lakhina2008}. The energetic ions are described by a nonthermal
or Cairns {\it et al.} distribution \cite{Cairns1995,Mamun1996,Mamun2002}, and are referred
to as nonthermal ions. The latter is now being common feature of the geospace plasmas,
and in general it is turning out to be a characteristic feature of space plasmas \cite{Futaana2003}.
For examples, nonthermal ions are observed in the Earth's bow-shock region by the Vela satellite \cite{Asbridge1968}; ASPERA on the Phobos 2 satellite has observed the loss of energetic ions
from the upper ionosphere of Mars; in solar active region \cite{Imada2009}.

In this paper, we study the effect of non-thermal (energetic or fast) ions on small and
large amplitude EA solitary pulses (EASPs) and EA-DLs in a multi-component plasma.
It is found that the presence of non-isothermal ions greatly affect the features of
both EASPs and EA-DLs. The latter can be identified as localized
electrostatic excitations in observational data from space and laboratory plasmas.

\section{Governing Equations}

We consider the nonlinear propagation of the EAWs in one-dimensional, collisionless, unmagnetized plasmas composed
of cold electrons, hot electrons obeying a Boltzmann
distribution, and nonthermal ions following nonthermal
distribution. Thus, at equilibrium, we have $n_{c0} +
n_{h0}=n_{i0}$, where $n_{i0}$ is the nonthermal ion number
density at equilibrium. The nonlinear dynamics of the EAWs
propagating in such a plasma system is governed by

\begin{eqnarray}
&&\frac{\partial n_c}{\partial t}+\frac{\partial}{\partial
x}(n_cu_c)=0,\label{c}\\
&&\frac{\partial u_c}{\partial t}+u_c\frac{\partial u_c}{\partial
x}=\frac{\partial \Psi}{\partial x},\label{m}\\
&&\hspace*{-8mm}\frac{\partial^2 \Psi}{\partial x^2}=\mu
e^\Psi+(1-\mu)n_c-(1+\beta \sigma \Psi+\beta
\sigma^2\Psi^2)e^{-\sigma \Psi},\label{p}
\end{eqnarray}
where $n_c$ is the cold electron number density normalized by its
equilibrium value $n_{c0}$, $u_c$  the cold electron speed
normalized by $C_e$, $\Psi$ the wave potential normalized by
$k_BT_h/e$, $\mu=n_{h0}/n_{i0}$, $\sigma=T_h/T_i$, $T_i$  the
ion temperature, and $\beta=4\alpha/(1+3\alpha)$, in which
$\alpha$ is the nonthermal parameter
\cite{Cairns1995,Mamun1996,Mamun2002}. The time variable $t$ is
in units of  $\omega^{-1}_{pc}$, and the space variable $x$ is
normalized by $\lambda_{dh}$. Our three-component plasma model is
valid for electrostatic disturbances with the EAW phase speed much larger
(smaller) than the thermal speed of the cold electrons (hot electrons and hot ions).
The non-isothermal ion density distribution is associated with an ion
distribution function that departs from the Maxwell-Boltzmann law
on account of the EA wave amplitude dependent of the ion energy
in the nonlinear regime.

\section{Derivation of Energy Integral}

To derive an energy integral \cite{Bernstein1957,Sagdeev1966}
from Eqs. (\ref{c})-(\ref{p}), we first make all the dependent
variables to depend only on a single variable $\xi=x-Mt$, where $M$
is the nonlinear wave speed normalized by $C_e$. We note that $M$
is not the Mach number, since it is normalized by $C_e$, which is
not the EAW phase speed. Now, using the steady-state condition and
imposing the appropriate boundary conditions (namely,
$n_c\rightarrow1$, $u_c\rightarrow0$, and $\Psi\rightarrow0$ at
$\xi\rightarrow\pm\infty$, one can express $n_c$ as

\begin{eqnarray}
&&n_c=\left(1+\frac{2\Psi}{M^2}\right)^{-1/2}. \label{nc}
\end{eqnarray}
Now, substituting Eq. (\ref{nc}) into Eq. (\ref{p}), multiplying
the resulting equation by $d\Psi/d\xi$, and applying the boundary
condition, $d\Psi/d\xi\rightarrow0$ at $\xi\rightarrow\pm\infty$,
one obtains an energy integral

\begin{eqnarray}
&&\frac{1}{2}\left(\frac{d\Psi}{d\xi}\right)^2+V(\Psi)=0,
\label{e}
\end{eqnarray}
for an oscillating particle of unit mass, with pseudo-position
$\Psi$, pseudo-time $\xi$, and a pseudo-potential $V(\Psi)$.
The latter for our purposes reads

\begin{eqnarray}
&&V(\Psi)=\mu(1-e^\Psi)+(1-\mu)M^2\left(1-\sqrt{1+\frac{2\Psi}{M^2}}\right)\nonumber\\
&&-\frac{e^{-\sigma\Psi}}{\sigma}\left(1+3\beta+3\beta\sigma\Psi+\beta\sigma^2\Psi^2\right)+\frac{1}{\sigma}(1+3\beta),
\label{v}
\end{eqnarray}
which is valid for the arbitrary amplitude EASPs and EA-DLs in our plasma.

\section{Small Amplitude Results}

We first investigate the properties of small amplitude
EASPs and EA-DLs by using a pseudo-potential approach
\cite{Bernstein1957,Sagdeev1966}. The expansion of $V(\Psi)$
around $\Psi=0$ is

\begin{eqnarray}
&&V(\Psi)=C_2\Psi^2+C_3\Psi^3+C_4\Psi^4+\cdot\cdot\cdot\cdot,\label{v1}
\end{eqnarray}
where

\begin{eqnarray}
&&C_2=\frac{1}{2}\left[\frac{1-\mu}{M^2}-\mu-\sigma(1-\beta)\right],\label{C2}\\
&&C_3=\frac{1}{6}\left[-\frac{3(1-\mu)}{M^4}-\mu+\sigma^2\right],\label{C3}\\
&&C_4=\frac{1}{24}\left[\frac{15(1-\mu)}{M^6}-\mu-\sigma^3(1+3\beta)\right].\label{C4}
\end{eqnarray}

\subsection{EA Solitary Pulses}

Let us first consider small-amplitude EASPs for which
$V(\Psi)=C_2\Psi^2+C_3\Psi^3+C_4\Psi^4$ holds. This
approximation allows us to write the small-amplitude solitary
wave solution \cite{Mannan2011} of Eq. (\ref{e}) as

\begin{eqnarray}
\Psi=\left[\frac{1}{\Psi_{m2}}-\left(\frac{1}{\Psi_{m2}}
-\frac{1}{\Psi_{m1}}\right)\cosh^2\left(\frac{\xi}{\delta}\right)\right]^{-1},
\label{ssol}
\end{eqnarray}
where

\begin{eqnarray}
&&\Psi_{m1,2}=\Psi_m\left[1\mp\sqrt{1-\frac{C_2}{C_0}}\right],\nonumber\\
&&\Psi_m=-\frac{C_3}{2C_4},\nonumber\\
&&\delta=\sqrt{-\frac{2}{C_4\Psi_{m1}\Psi_{m2}}},\nonumber\
\end{eqnarray}
$\delta$ is the width of the EASPs, and $C_0=C_3^2/4C_4$.
\begin{figure}[t!]
\centerline{\includegraphics[width=8cm]{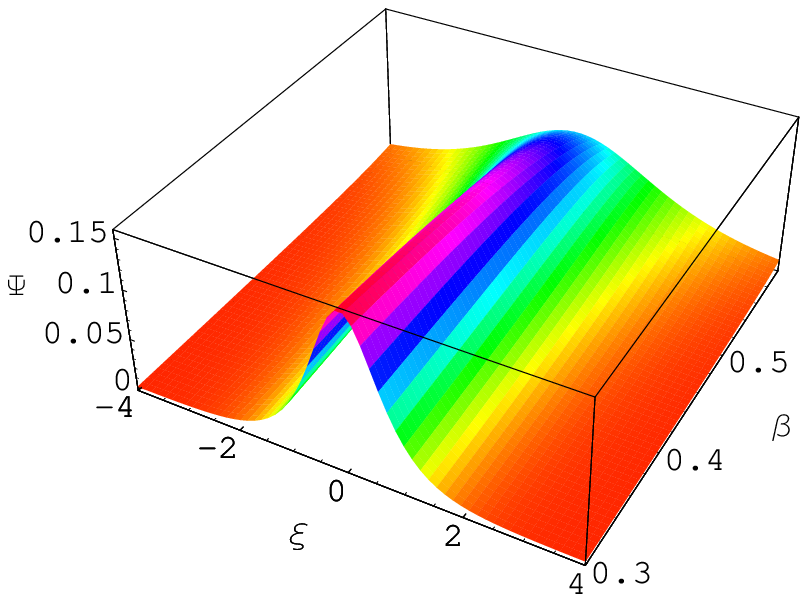}} \caption{(Color
online) Showing how the amplitude and the width of the positive
EASPs vary with $\beta$ for $\sigma=10$, $\mu=0.5$, and $M=0.36$.}
\label{sp}
\end{figure}
\begin{figure}[t!]
\centerline{\includegraphics[width=8cm]{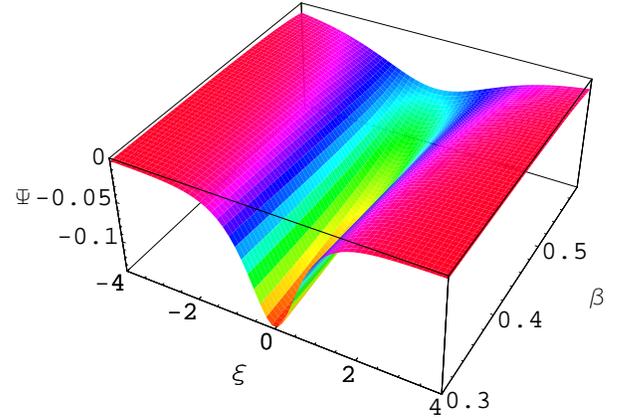}} \caption{(Color
online) Showing how the amplitude and the width of the negative
EASPs vary with $\beta$ for $\sigma=10$, $\mu=0.5$, and $M=0.35$.}
\label{sn}
\end{figure}

The profiles (indicating the amplitude and width) of the small
amplitude EASPs associated with positive and negative potential
are graphically displayed in figures \ref{sp} and \ref{sn}. It is
seen from figure \ref{sp} that the amplitude (width) of the
positive EASPs decreases (increases) with $\beta$. On the other
hand, figure \ref{sn} reveals that the amplitude (width)
of the negative EASPs decreases (increases) with $\beta$.

\subsection{EA Double Layers}

To study the small but finite amplitude EA-DLs,
$V(\Psi)=C_2\Psi^2+C_3\Psi^3+C_4\Psi^4$ allows us to write the
double layer solution \cite{Mamun2011} of Eq. (\ref{e}) as

\begin{eqnarray}
\Psi=\frac{\Psi_m}{2}\left[1+\tanh\left(\frac{\xi}{\Delta}\right)\right],
\label{dlsol}
\end{eqnarray}
where

\begin{eqnarray}
&&\Psi_m=-\frac{C_3}{2C_4}=-\frac{2C_2}{C_3},\nonumber\\
&&\delta=\frac{2}{\Psi_m\sqrt{-2C_4}}.\nonumber\
\end{eqnarray}
\begin{figure}[t!]
\centerline{\includegraphics[width=8cm]{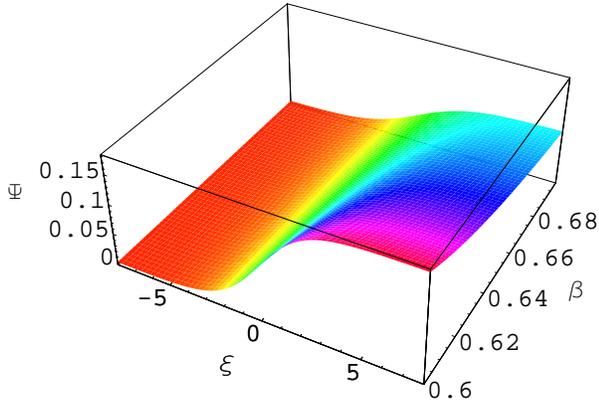}} \caption{(Color
online) Showing how the amplitude and the width of the positive
EA-DLs vary with $\beta$ for $\sigma=10$, $\mu=0.5$, and $M=0.38$.}
\label{dl}
\end{figure}
The profiles (indicating height and thickness) of the small
amplitude EA-DLs associated with positive potential are
graphically displayed in figure \ref{dl}. It is seen from figure
\ref{dl} that the amplitude (width) of the EA-DLs decreases
(increases) with $\beta$.

\section{Arbitrary Amplitude EA Solutions}

We now investigate the properties of arbitrary amplitude
EASPs and EA-DLs by numerical analyses of the
pseudo-potential $V(\Psi)$, given by (\ref{e}). It is evident from
(\ref{e}) that $V(\Psi)=dV(\Psi)/d\Psi=0$ at $\Psi=0$. Therefore,
the EASP and EA-DL solutions of (\ref{e}) exist if
$d^2V(\Psi)/d\Psi^2)_{\Psi=0}<0$ i.e., $C_2>0$, i.e.,
$M>M_c=\sqrt{(1-\mu)/(\mu+\sigma-\beta\sigma)}$, so that the
fixed point at the origin is unstable.
\begin{figure}[t!]
\centerline{\includegraphics[width=8cm]{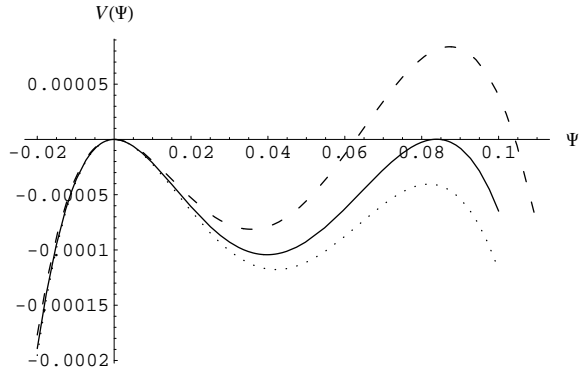}} \caption{The
profile  of the pseudo-potential for $\mu=0.5$, $\sigma=10$,
$\beta=0.7$, $M=0.4133$ (solid curve), $M=0.41$ (dash curve), and
$M=0.415$ (dotted curve).} \label{a1}
\end{figure}
\begin{figure}[t!]
\centerline{\includegraphics[width=8cm]{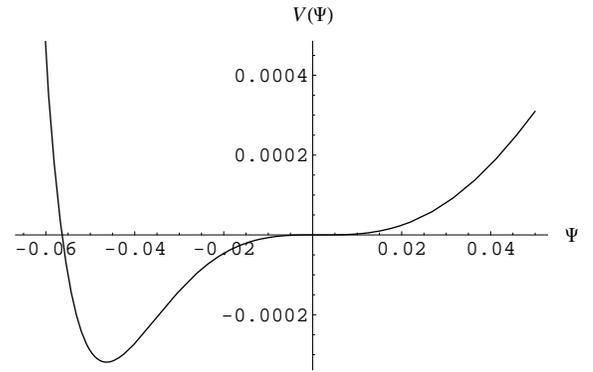}} \caption{The
profile  of the pseudo-potential for $\mu=0.5$, $\sigma=10$,
$\beta=0.7$, and $M=0.38$.} \label{a2}
\end{figure}

One can easily show by numerical analyses of $V(\Psi)$ [given
in (\ref{e})] that the EASPs exist with both positive potential
($\Psi>0$) and negative potential ($\Psi<0$), but the EA-DLs exist
only with positive potential ($\Psi>0$). A part of the numerical
analyses, showing the formation of the potential wells in the
positive $\Psi$-axis, i.e. showing the existence of the EASPs
and EA-DLs with $\Psi>0$, is displayed in Fig. \ref{a1}. Figure
\ref{a1} displays the formation of the potential wells in the
positive $\Psi$-axis, which corresponds to the formation of the
EASPs and EA-DLs with positive potential for $\mu=0.5$,
$\sigma=10$, and $\beta=0.7$. The Mach numbers are $0.4133$
(solid curve), corresponding to a double layer solution, $0.41$
(dash curve), yielding a positive solitary solution, $0.415$
(dotted curve), giving no positive solitary structure. Figure
\ref{a2} exhibits the formation of the potential well in the
negative $\Psi$-axis, which corresponds to the formation of
EASPs with negative potential for $\mu=0.5$, $\sigma=10$,
$\beta=0.7$, and $M=0.38$ (solid curve).
\begin{figure}[t!]
\centerline{\includegraphics[width=8cm]{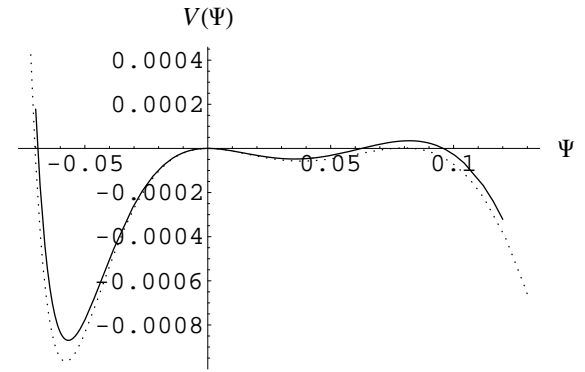}} \caption{The
formation of potential wells in both positive and negative
$\Psi$-axes at $\mu=0.3$, $\sigma=10$, $\beta=0.583$, $M=0.41$
(solid curve), and $M=0.411$ (dotted curve).} \label{a3}
\end{figure}
\begin{figure}[t!]
\centerline{\includegraphics[width=8cm]{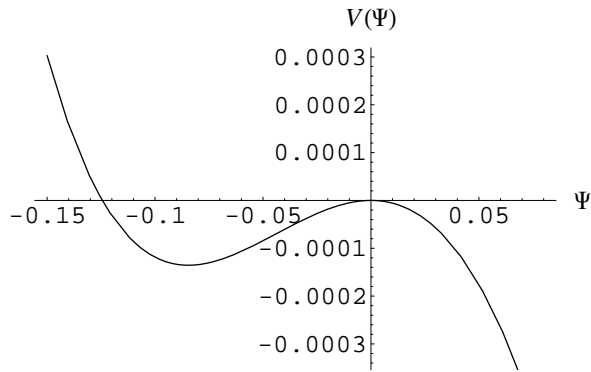}} \caption{The
formation of potential well in negative $\Psi$-axes at $M=0.95$,
$\mu=0.5$, $\sigma=0.4$, and $\beta=0.6$.} \label{a4}
\end{figure}

Figure \ref{a3} exhibits that for $M>M_c$ positive and negative
EASPs coexist (solid curve), and EA-DLs are formed with positive
potential only, but not with negative potential. It also shows
that negative EASPs  and positive EA-DLs coexist (dotted curve).
Figure \ref{a4} depicts the formation of the potential well in the
negative $\Psi$-axis, which corresponds to the formation of
EASPs with negative potential for $\mu=0.5$, $\sigma=0.4$,
$\beta=0.6$, and $M=0.95$. Figure \ref{a1} can provide a
visualization of the amplitude ($\Psi_m$) and the width
[$|\Psi_m|/|V_m|$, where $|V_m|$ is the minimum value of $V
(\Psi)$ in the potential wells formed in the positive $\Psi$
-axis]. Figure \ref{a2} can also provide a visualization of the
height and the thickness in the potential well formed in the
negative $\Psi$ -axis]. Figure \ref{a1}, where the values of $M$
are around its critical value ($M_c$), indicate the existence of
arbitrary amplitude EASPs and EA-DLs. Figure \ref{a2}, where the
values of $M$ are around its critical value ($M_c$), reveals the
existence of arbitrary amplitudes EASPs. It is found from this
visualization (after a more numerical analysis with different
values of $\mu$, $\sigma$, and $\beta$, which are not shown here)
that the variation of the amplitude and the width with $\mu$,
$\sigma$, and $\beta$ in the case of arbitrary amplitude EASPs
and EA-DLs are exactly the same as that in the case of small
amplitude EASPs and EA-DLs.

\section{Summary and Conclusions}

In this paper, we have considered a plasma composed of cold inertial electrons,
hot Boltzmann distributed electrons, and non-isothermal hot ions, and have investigated
properties of small but finite, as well as arbitrary amplitudes EASPs and EA-DLs.
We have used the pseudo-potential approach, which is valid for arbitrary amplitudes
EASPs and EA-DLs. It has been found for the small amplitude limit that (i)
the non-thermal plasma system under consideration is found to
support EASPs and EA-DLs, whose salient features
(the amplitude, the width, the speed, etc.) are significantly modified by the
non-thermal parameter $\beta$; (ii) the amplitude (width) of the
EASPs decreases (increases) with $\beta$; (iii) the amplitude
(width) of the EA-DLs decreases (increases) with $\beta$. On the
other hand, it has been found for the arbitrary amplitude that (i)
EA-DLs are formed at $M=0.4133$; (ii) positive EASPs are formed
at $M=0.41$; (iii) no positive EASPs are formed at $M=0.415$;
(iv) the negative EASPs are formed at $M=0.38$; (v) at $M>M_c$
the negative EASPs and positive EASPs or EA-DLs coexist; (vi)
when ion temperature is greater than hot ion temperature i.e.
$\sigma=0.4$ then we get only negative EASPs.

The ranges of different plasma parameters used in our
investigation are very wide ($\mu=0.2-0.9$, $\sigma=0.4-10$, and
$\beta=0.3-0.8$), are relevant to both space
\cite{Dubouloz1991,Dubouloz1993,Pottelette1999,Berthomier2000,Singh2001}
and laboratory plasmas \cite{Derfler1969,Henry1972}. Thus, the
results of the present investigation should help us to explain
salient features of localized EA perturbations propagating in
space and laboratory plasmas that have two distinct groups of electrons
and a component of non-isothermal hot ions.
\acknowledgements
The research grant for research equipment from the Academy of World Sciences (TWAS), Trieste,
Italy is gratefully acknowledged.

\end{sloppypar}
\end{document}